**Information Seeking Responses To News Of Local COVID-19 Cases: Evidence From Internet Search Data**

**Ana I. Bento[1]\*, Thuy Nguyen[2], Coady Wing[2], Felipe Lozano-Rojas[2], Yong-Yeol Ahn[3], Kosali Simon[2]\***

1 School of Public Health, Indiana University, Bloomington

2 O'Neill School of Public and Environmental Affairs, Indiana University, Bloomington

3 Luddy School of Informatics, Computing and Engineering, Indiana University, Bloomington

**\*Corresponding authors**

**Abstract**

The novel coronavirus (COVID-19) outbreak is a global pandemic with community circulation in many countries, including the U.S. where every state is reporting confirmed cases. The course of this pandemic will be largely shaped by how governments enact timely policies, disseminate the information, and most importantly, how the public reacts to them. Here, we examine information-seeking responses to the first COVID-19 case public announcement in a state. By using an event-study framework, we show that such news increases collective attention to the crisis right away, but the elevated level of attention is short-lived, even though the initial announcements were followed by increasingly strong measures.

We find that people respond to the first report of COVID-19 in their state by immediately seeking information about COVID-19, as measured by searches for coronavirus, coronavirus symptoms and hand sanitizer. On the other hand, searches for information regarding community level policies (e.g., quarantine, school closures, testing), or personal health strategies (e.g., masks, grocery delivery, over-the-counter medications) do not appear to be immediately triggered by first reports. These results are encouraging given our study period is relatively early in the epidemic and more elaborate policy responses were not yet part of the public discourse. Further analysis will track evolving patterns of responses to subsequent flows of public information.

**Introduction**

The first confirmed case of COVID-19 in the U.S. occurred in Washington State on January 21 2020. Since then the virus has spread across the country (1,2). There are confirmed cases in every state and the vast majority are not connected to international travel (3), indicating that the virus has been circulating for several weeks before the first positive test (4,5). Slowing the transmission of the virus will help reduce the burden of the disease, save lives, and reduce strain on the health care system (4,6). In the absence of a vaccine, the main strategies for reducing transmission involve sanitation and handwashing, disciplined social distancing, quarantines, and school and workplace closures (6). These non-pharmaceutical interventions (NPIs) are not disease-specific, yet they are effective and powerful ways to control transmission (7), as shown by earlier epidemics (8) and the steady decline of COVID-19 cases in China (2). To be successful, NPIs require people to voluntarily undertake behavioral changes that may be personally costly. This is particularly true in countries with deeply rooted norms about personal freedom and reluctance to impose mandatory policies.

In settings where restrictions on the free movement of residents are less strictly enforced, it is especially critical for public officials to know whether and for how long statements from leaders motivate individuals to efficiently seek and absorb specific information as officials put their communication strategies in place. Such coordinated actions are needed at the first signs of community spreading and cannot be guided through traditional polling methods. Because the extent of the epidemic is likely to be underestimated at this stage, it is crucial to know how much the early case announcements induced heightened collective attention and information-seeking behavior.

To provide rapid information to guide policy making, we use Internet search data in an event-study design to examine how collective attention and information seeking behaviors respond to state government announcements of first COVID-19 cases. We highlight changes in search patterns occurring in the days leading up to and following first case announcements in a state (1,2).

Results

Our results display Poisson coefficients and confidence intervals for the 40 days leading up to the first case announcement, and up to the 20 days following it. The vertical axis measures the differential percentage change in search frequency as states approach the first case announcement.

In Figure 1 depicts the event study estimates of the effects of initial announcements on overall coronavirus search. Searches for "coronavirus" increased by about 36% (95% CI: 27% to 44%) on the day immediately after the announcement but quickly decreased back to the baseline level in less than a week or two (Figure 1). There was no observable trend in the search behavior in the days leading up to the announcements, suggesting the first "local" case indeed heightened the collective attention to the pandemic. However, the increased level of information-seeking faded within two weeks, even though many announcements of school closures or other mitigation strategies followed, suggesting that increased attention is only short-lived. This is consistent with no observable trend in the sense of urgency prior to the local announcement.

Figure 2 shows the event studies of Internet search for (i) symptoms and treatments, (ii) hand sanitizer and diagnostic tests, (iii) coordinated responses, and (iv) narratives that undermine public responses. News of the first COVID-19 case in a state leads to a 52% increase in searches for "coronavirus symptoms" but did not increase searches for coronavirus treatment options. The second row shows searches for "hand sanitizer" increasing 35% immediately after first case announcements, and unlike in the previous two cases, the search activity remains high for the remainder of the observable period. However, the announcements did not induce searches for nearby coronavirus testing opportunities, at least in the period under study. The third row suggests that first case announcements did not induce search for community level policies (quarantines, school closures, and coronavirus testing), or more elaborate personal health strategies (face masks, grocery delivery, over-the-counter medications). The final row examines how first case reports affect searches about the credibility of the epidemic. There is no indication that government confirmation of the first case increased or reduced searches for coronavirus hoaxes and overreactions; one might have expected official news to reduce concerns about false news.

Conclusion

Our results suggest that first state COVID-19 case announcements do lead to a widespread increase in the extent to which people seek out Internet information about the epidemic. We find that announcements of first cases had the biggest effect on searches for basic information about the virus and its symptoms, suggesting that people attempted to educate themselves about COVID-19 and its effects. However, first case reports do not lead to a differential increase in searches involving large scale non-pharmaceutical interventions (school and workplace closures) that will likely have important consequences for everyday life. This may be because first cases are reported relatively early in the epidemic and many of these more serious mitigation strategies were not yet part of public discourse. Finally, although one might expect official news to reduce concerns related to false messages, we do not find much indication that state announcements of first COVID-19 cases affect searches questioning whether the epidemic may be a hoax. Overall, our analysis of Internet search data suggests that government information disclosure does help focus public attention on the crisis. Our evidence also indicates people seem to mainly react by seeking information on what they can and should do in response to the epidemic. This provides actionable evidence for those designing the next stages of public information strategies for the COVID-19 crisis and future threats to public wellbeing.

Materials and Methods

Samples. Our analysis is based on a balanced daily panel of COVID-19 related search intensity data in 50 US states and D.C. between January 1, 2020 and March 18, 2020 (n=51 regions x 78 days). We collected the data using a restricted-access Google Health Trends Application Program Interface (API) account (12).

Measures. The outcome variables measure the daily share of all Google queries in a state that correspond to a particular term during our study period. We multiply the shares by 10 million and round to the nearest integer to make the measure more interpretable. We collected data on the timing of the first COVID-19 case announcements from media reports in each state; see Supplement Table 1. All states reported their first case by March 17, 2020.

Data Analysis. We estimated Poisson models in an event-study (13) framework to examine the effect of the first COVID-19 case announcements on searches. In the model we use, the expected number of searches per 10 million is:

$$y_{i,t} = e^{(\beta_0 + \sum_{k=[-41,31]} \beta_k Announcement_{i,k} + \theta_i + \theta_t + \varepsilon_{it})}$$

Where $Announcement_{i,k}$ is an indicator variable set to 1 if the first case of state i was announced k days ago. The reference category was one day prior to the first case announcement and we created single indicators $Announcement_{i,31}$, for any $k \geq 30$ and $Announcement_{i,-41}$, for any $k \leq -40$.. We controlled for state fixed effects to allow for time invariant differences in search patterns by state, and for date fixed effects to account national trends. Standard errors allowed for heteroskedasticity and clustering at the state level. We performed regression analysis using Stata, version 16.0 (StataCorp).

Data availability. State news data and code are available here (14). The search data for this study are available from Google Trends, but are restricted in use; researchers may apply to Google Trends API for access.

Limitations. Although the eventual failure of Google Flu Trends suggested that building and maintaining a complex model for a long period of time is difficult (9), internet search volume has been shown to be a good proxy for many socio-economic behavioral indicators, such as automobile sales (11) or dietary patterns (12). Moreover, the major issues with search query data — long-term drift and representation bias across geographic regions — does not come into play in our study because we focus on relative change in a short time window surrounding the government announcements.

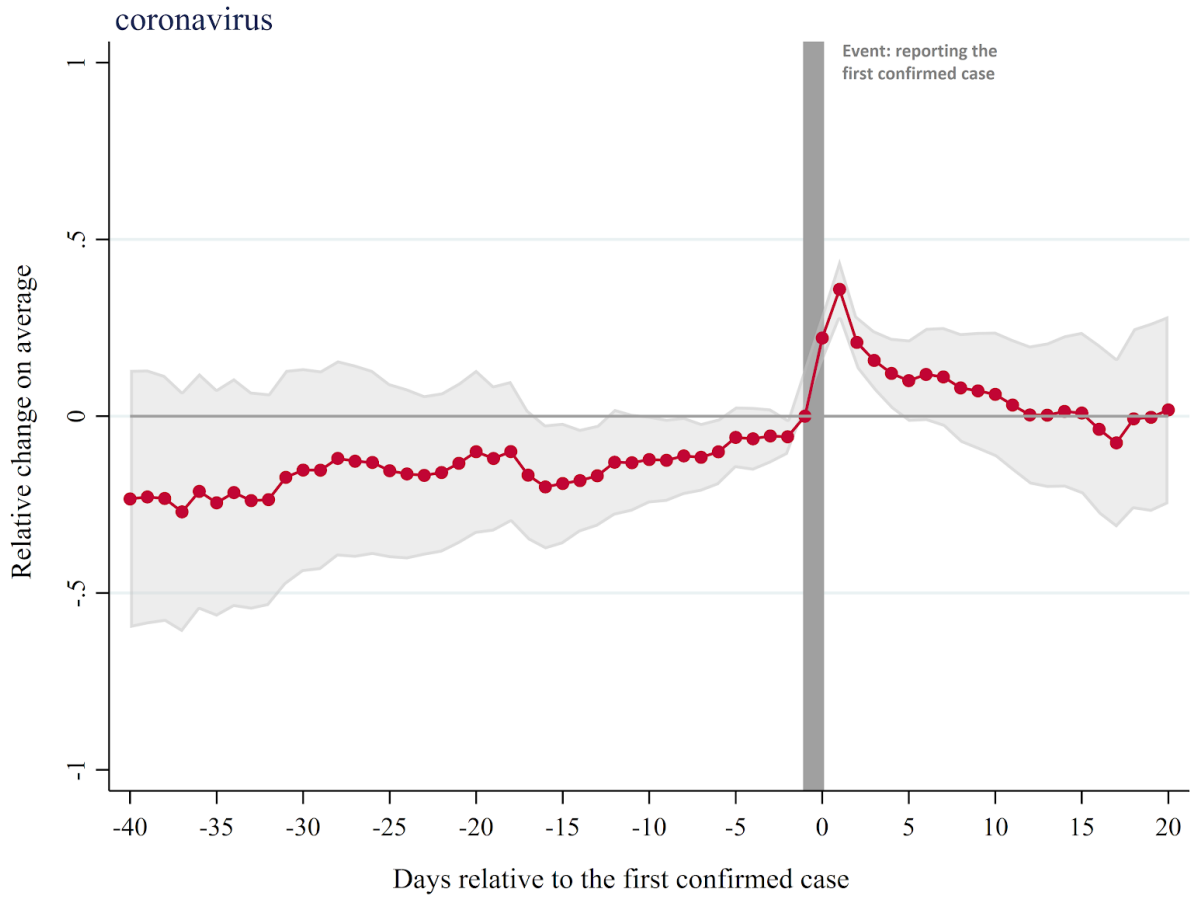

Figure 1: Time varying effects of announcements of the first COVID-19 case in a state on searches for coronavirus. The period prior to the treatment (first confirmed case) is set as a reference - grey vertical bar. In red are the estimated coefficients (95% CI grey band) in the Poisson model (differences in log-expected counts of search relative to the period prior to the event). The average search frequency of this term is 97,023.9 per state per day.

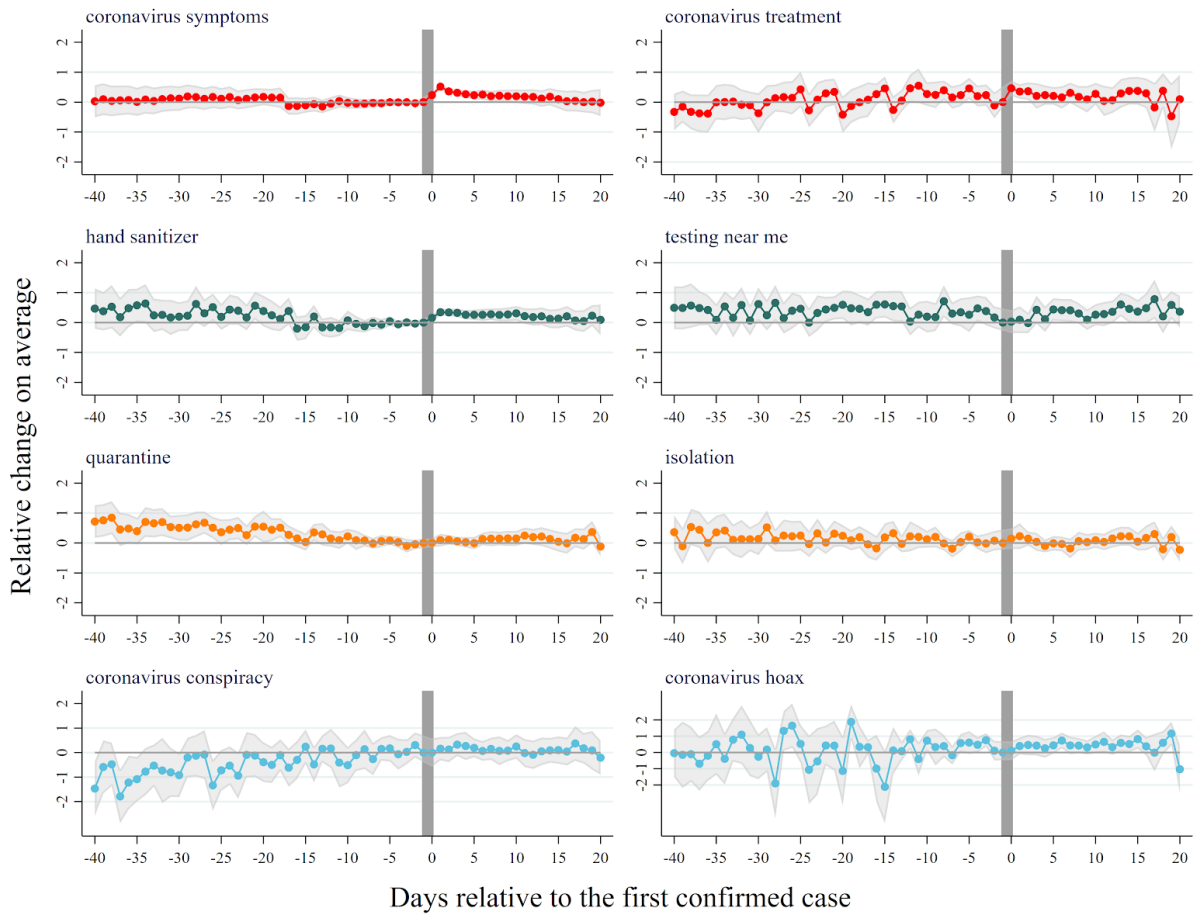

Figure 2. Time varying effects of announcements of the first COVID-19 case in a state on searches for (i) symptoms and treatments (red), (ii) hand sanitizer and diagnostic tests (green), (iii) coordinated responses (orange), and (iv) narratives that undermine public responses (blue). Point estimated coefficients are the dotted lines (95% CI grey band) in the Poisson models (differences in log-expected counts of search relative to the period prior to the event). Average search frequency per state per day for the terms: coronavirus symptoms (8,522.6), coronavirus treatment (448.2), hand sanitizer (3,214.7), testing near me (381.6), quarantine (2,158.2), isolation (631.2), coronavirus conspiracy (205.1), coronavirus hoax (123.0).

**Acknowledgements:** Raj Thakkar, Chrislin Priscilla for research assistance

**Author contributions:** FLR and KS collected the data; FLR, TN, KS, AIB, and CW analyzed data; and AIB, FLR, TN, KS, YYA, and CW contributed to the writing of the paper. The authors declare no competing interest.